# A Simple Standard for Sharing Ontological Mappings (SSSOM)

Nicolas Matentzoglu[1], James P. Balhoff[2], Susan M. Bello[3], Chris Bizon[2], Matthew Brush[4], Tiffany J. Callahan[4], Christopher G Chute[5], William D. Duncan[6], Chris T. Evelo[7], Davera Gabriel[5], John Graybeal[8], Alasdair Gray[9], Benjamin M. Gyori[10], Melissa Haendel[4], Henriette Harmse[11], Nomi L. Harris[6], Ian Harrow[12], Harshad Hegde[6], Amelia L. Hoyt[13], Charles T. Hoyt[10], Dazhi Jiao[5], Ernesto Jiménez-Ruiz[14,15], Simon Jupp[16], Hyeongsik Kim[17], Sebastian Koehler[18], Thomas Liener[12], Qinqin Long[19,] James Malone[20], James A. McLaughlin[11], Julie A. McMurry[4], Sierra Moxon[6], Monica C. Munoz-Torres[4], David Osumi-Sutherland[11], James A. Overton[21], Bjoern Peters[22], Tim Putman[4], Núria Queralt-Rosinach[19], Kent Shefchek[4], Harold Solbrig[5], Anne Thessen[4], Tania Tudorache[23], Nicole Vasilevsky[4], Alex H. Wagner[24,25], Christopher J. Mungall[6,*]

\* Corresponding author, cjmungall@lbl.gov

1 Semanticly Ltd., London WC2H 9JQ, UK
2 RENCI, University of North Carolina, Chapel Hill, NC 27517, USA
3 The Jackson Laboratory, Bar Harbor, ME 04609, USA
4 University of Colorado Anschutz Medical Campus, Aurora, CO 80217, USA
5 Johns Hopkins University, Baltimore, MD 21210, USA
6 Lawrence Berkeley National Laboratory, Berkeley, CA 94720, USA
7 Maastricht University, The Netherlands
8 Stanford University, Stanford, CA, 94305, USA
9 Department of Computer Science, Heriot-Watt University, Edinburgh, UK
10 Harvard Medical School, Boston, MA 02115, USA
11 European Bioinformatics Institute (EMBL-EBI), Hinxton, UK
12 Pistoia Alliance Inc, USA
13 Beth Israel Deaconess Medical Center, Boston, MA, USA
14 City University of London, UK
15 University of Oslo, Norway
16 SciBite Limited, Bio Data Innovation Centre, Wellcome Genome Campus, Hinxton, Saffron Walden CB10 1DR, UK
17 Robert Bosch LLC
18 Ada Health GmbH
19 Leiden University Medical Center, Leiden, The Netherlands
20 BenchSci, 25 York St Suite 1100, Toronto, ON M5J 2V5, Canada
21 Knocean Inc., Toronto, Ontario, Canada
22 La Jolla Institute for Immunology, 9420 Athena Circle, La Jolla, CA 92037
23 Independent Scholar
24 The Steve and Cindy Rasmussen Institute for Genomic Medicine, Nationwide Children's Hospital, Columbus, OH 43205
25 The Ohio State University College of Medicine, Columbus, OH 43210, USA

## Abstract

Despite progress in the development of standards for describing and exchanging scientific information, the lack of easy-to-use standards for mapping between different representations of the same or similar objects in different databases poses a major impediment to data integration and interoperability. Mappings often lack the metadata needed to be correctly interpreted and applied. For example, are two terms equivalent or merely related? Are they narrow or broad matches? Or are they associated in some other way? Such relationships between the mapped terms are often not documented, which leads to incorrect assumptions and makes them hard to use in scenarios that require a high degree of precision (such as diagnostics or risk prediction). Furthermore, the lack of descriptions of how mappings were done makes it hard to combine and reconcile mappings, particularly curated and automated ones.

We have developed the Simple Standard for Sharing Ontological Mappings (SSSOM) which addresses these problems by:

1. Introducing a machine-readable and extensible vocabulary to describe metadata that makes imprecision, inaccuracy and incompleteness in mappings explicit.
2. Defining an easy to use simple table-based format that can be integrated into existing data science pipelines without the need to parse or query ontologies, and that integrates seamlessly with Linked Data standards.
3. Implementing open and community-driven collaborative workflows that are designed to evolve the standard continuously to address changing requirements and mapping practices.
4. Providing reference tools and software libraries for working with the standard.

In this paper, we present the SSSOM standard, describe several use cases in detail, and survey some of the existing work on standardizing the exchange of mappings, with the goal of making mappings Findable, Accessible, Interoperable, and Reusable (FAIR). The working draft of the SSSOM specification can be found at http://w3id.org/sssom/spec.

# Introduction

The problem of mapping between different identifiers is ubiquitous in bioinformatics, and more generally in data science and data management. The problem arises when the same or similar entity or concept is assigned different identifiers in different databases or vocabularies, necessitating the need to maintain mappings between these identifiers if information is to be combined. For example, a single gene, or a single disease entity such as Fanconi anemia may be assigned multiple different identifiers in different databases (Figure 1). If data from these databases is merged without mappings, then information related to the same entity, such as Fanconi anemia, is  combined, potentially losing crucial insights. Creating and maintaining mappings is costly, and the cost of incorrect or incomplete mappings can be even higher. For example, if health information is transferred between different systems, inaccurate mappings between disease terms could result in less accurate or even completely wrong diagnoses, with potentially serious negative consequences.

Despite the importance of the mapping problem, there is no single widely agreed-upon standard for exchanging mappings. Instead, various ad-hoc schemes and formats are typically used, and these schemes frequently omit crucial information (see Table 1). Many available mappings are just single-use conversion tables between two particular databases. These mappings generally have limitations: they are usually incomplete or inaccurate in ways that are non-transparent; they lack sufficient metadata to allow reuse in different contexts; and they do not follow FAIR (Findable, Accessible, Interoperable and Reusable) standards (1). Addressing these limitations is the central aim of the SSSOM standard.

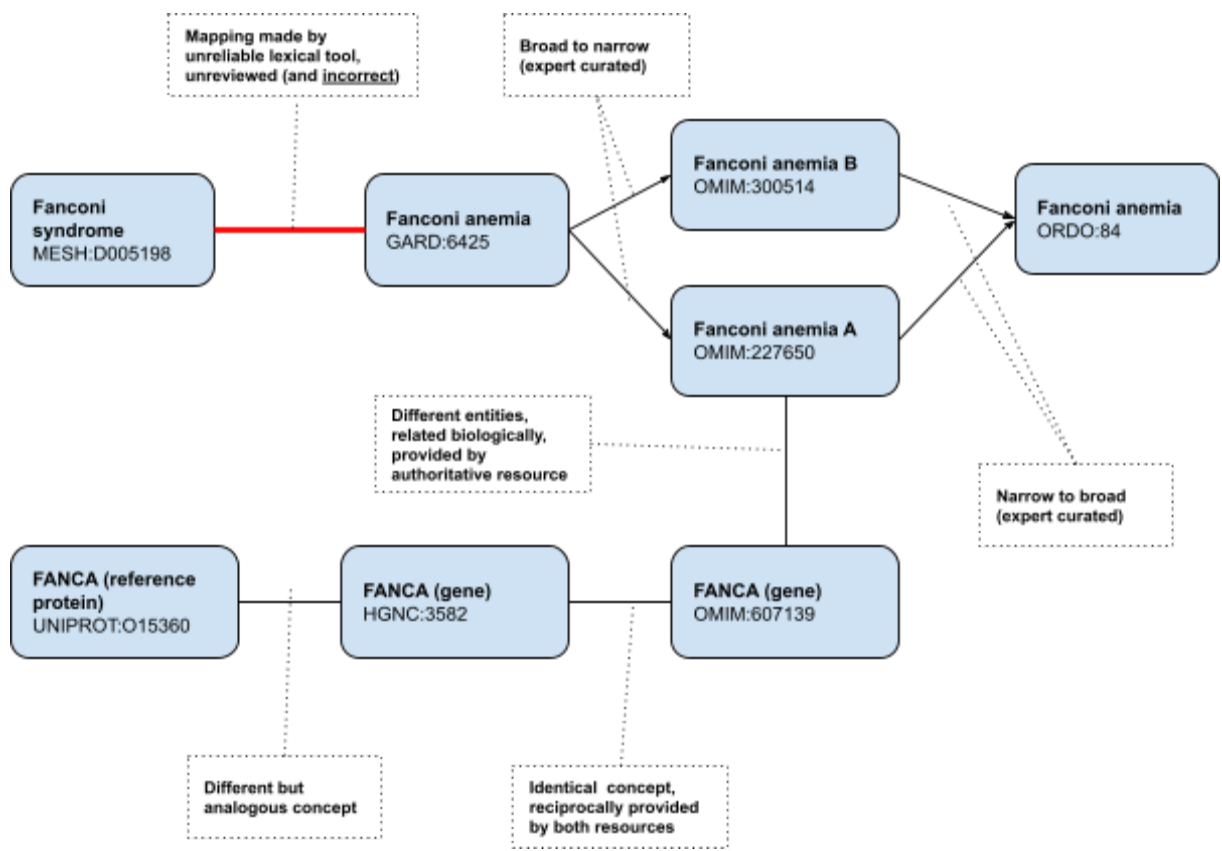

Figure 1: Example of mappings between different identifiers representing statements about similarity or identity of concepts across resources and vocabularies. Even with this simplified example, it is possible to see a range of mapping types, and that providing information about each mapping is crucial to understanding the bigger picture. This information helps avoid errors such as mistakenly conflating two variants of a disease.

## Desired features of a standard for mappings

We cataloged key characteristics of standard mappings, based on the diverse group of use cases described later in this paper. We usually refer to an entity that describes the relationship as the "predicate", but generally use the terms "predicate" and "relationship" loosely to mean the same thing.

| Feature | Why | Examples | Negative example |
| --- | --- | --- | --- |
| **Explicit relationship types** | Applications that demand highly accurate results require mapping relations with explicit precision and semantics | EC:2.2.1.2 exactMatch GO:0004801 (transaldolase activity) | Two-column file that maps FMA 'limb' to Uberon 'limb', hiding differences in species-specificity |
| **Explicit confidence** | Different use cases require different levels of confidence and accuracy | A mapping tool assigns a confidence score based on the amount of evidence that is explicitly recorded | Without the confidence score we cannot filter out automated mappings with low confidence |
| **Provenance** | Understanding how a mapping was created is crucial to interpreting it | Mapping file that automated mappings with link to tool used; curated mapping file with curators' ORCIDs provided | Two-column mapping file with no indication of how the mapping was made, and no supplementary metadata file |
| **Explicit declaration of completeness** | Must be able to distinguish between absence due to lack of information vs deliberate omission | Mapping file where rejected mappings are explicitly recorded | Mapping file where absence of a mapping can mean either explicitly rejected mapping OR the mapping was not considered/reviewed |
| **FAIR principles** | Mappings should be Findable, Accessible, Interoperable, and Reusable | Mapping file available on the web with clear licensing conditions, in standard format, with full metadata | Mapping files exchanged via email |
| **Unambiguous identifiers** | Mapping should make use of standard, globally | Standard ontology CURIEs like UBERON:0002101 for | Identifiers are used without explicitly defined prefixes; |
| | unambiguous identifiers such as CURIEs or IRIs | entities, with prefixes registered in a registry or as part of the metadata | mappings are created between strings rather than identifiers |
| **Allows composability** | Mappings from different sources should be combinable, and should be possible to chain mappings together | Defined mapping predicates (relations) such that reasoning about chains A->B->C is possible | Two mapping files with implicit or undefined relationships -> unclear whether these can be combined or composed |
| **Follows Linked Data standards** | Allows interoperation with semantic data tooling, facilitates data merging | All mapped entities have URIs, and metadata elements also have defined URIs; available in JSON-LD/RDF | No reuse of existing vocabularies for metadata or for relating mapped entities |
| **Well-described data model** | Allows interoperation and standard tooling | Data model provided in both human and machine readable form | Ad-hoc file format with unclear semantics |
| **Tabular representation** | Ease of curation and rapid analysis | A mapping available as a TSV that is directly usable in common data science frameworks; may complement a richer serialization | Ad-hoc flat file format requiring a custom parser |

**Table 1**: Desired features of a mapping standard, with examples of cases where the desired feature is met and examples where the desired feature is not met (negative examples).

These features include **explicitly declaring the relationship** between the two mapped entities. Frequently mappings are released as simple two-column files with no information about how the entities are related. Many applications benefit from or require mappings to be categorized as to whether the mappings are exact, or whether one concept is more general than the other, versus being closely related, but neither exact nor broader/narrower. There are a variety of different vocabularies that can be used to describe the relationship, including the Simple Knowledge Organization System (SKOS) (2) and the Web Ontology Language (OWL) (3), with different use cases dictating which system is used.

Additional desirable characteristics include various pieces of metadata associated with either a mapping collection or individual mappings, describing the **provenance** of the mapping (who made it, what tool made it if automated, when it was made), **versioning**, indications of **confidence** and **completeness**. This information helps humans understand and interpret the mappings, and can also be used by software.

We also include in our list of desiderata adherence to **FAIR principles** (1) and **Linked Data** standards (4). Linked Data standards aim to make data interoperable through the use of common data formats such as RDF and Uniform Resource Identifiers (URIs) for naming and identifying individual things. This includes making mappings easily available on the web as well as using standard URIs for representing both mapped entities and mapping data elements. There should also be a well-defined data model. Additionally, there should be a simple **tabular form** to enable easy exploration, management, viewing and processing by computational and human users, without needing specialized editing tools.

*Our solution.* In this paper, we present SSSOM, a Simple Standard for Sharing Ontological Mappings (pronounced "sessom" or S.S.S.O.M). SSSOM's goals are:

1. Providing a rich and easily extensible vocabulary for describing mapping metadata to address the aforementioned issues by encouraging the publication of mappings that are transparently imprecise, transparently inaccurate and transparently incomplete, as well as FAIR.
2. Offering a simple tabular format for the dissemination of mappings that can be easily integrated in typical data science toolchains.
3. Supporting a community-driven standard with well-defined governance and sustainable collaborative workflows.
4. Representing many different kinds of mappings, such as mappings between data models and their values, including literal values, controlled vocabularies and database entities.

## SSSOM: A rich and extensible vocabulary and schema for mapping metadata

In this section we describe the SSSOM standard in four subsections:

- The core data model and the metadata elements included in SSSOM
- How SSSOM is exchanged, including the canonical simple tabular serialization
- Governance and sustainability of the standard
- The emerging software ecosystem for working with SSSOM mappings

The complete SSSOM documentation and specification can always be retrieved via a permanent URL using the w3id system, https://w3id.org/sssom/ (5), and project information and source schema files can be found in our GitHub repository (https://github.com/mapping-commons/sssom). The current version of SSSOM at the time of this writing is $0.9$ (6). A detailed description can be found in the online documentation (5), but we will discuss many of the key features and their rationale later in this section.

At heart, the Simple Standard for Sharing Ontological Mappings (SSSOM) is a *simple* data model for representing mappings and mapping set metadata. "Simple" in this context means "flat", i.e. suitable for describing data that is primarily exchanged in tabular form such as TSV or CSV, as opposed to JSON, which allows for nested data structures. This simplicity, although it presents limitations (see section on Limitations), is one of the central design principles: the more complex structures like nested metadata or expressions (to represent entities) we allow, the more error-prone published mapping sets will become, and the more dependent we make users of the SSSOM standard on specific toolkits and software libraries - something we want to avoid as much as possible. Equally important, despite emerging toolkits for curating mappings, it is our experience that most mapping sets (certainly the ones used across all projects the authors are associated with) are curated as tables, which we will discuss later in this section. Despite this strong emphasis on simplicity, we are currently drafting a proposal to allow more deeply nested metadata (for example multiple mapping fields) and complex expressions (see Discussion section) through "profiles" that can be built on top of the current simple standard.

## Data model

The SSSOM data model describes individual pairwise mappings, which are grouped into mapping sets.

Each mapping can be described by up to 38 standard metadata "slots", or elements (in version 0.9). Four of these are required for any individual mapping: *subject_id, object_id* (the pair of entities mapped), *predicate_id* (the nature of the relationship between the two)*,* and *match_type* (how the mapping was derived). Additional optional metadata elements include *author_id*, *mapping_date*, and many more. For mapping sets, there are 23 elements, including *mapping_set_id*, *license* and *creator_id*.

All identifiers used in SSSOM should be CURIEs (7), i.e. prefixed identifiers with a registered prefix, following identifier best practice (8).

An example mapping with a few select metadata elements can be seen in Figure 2.

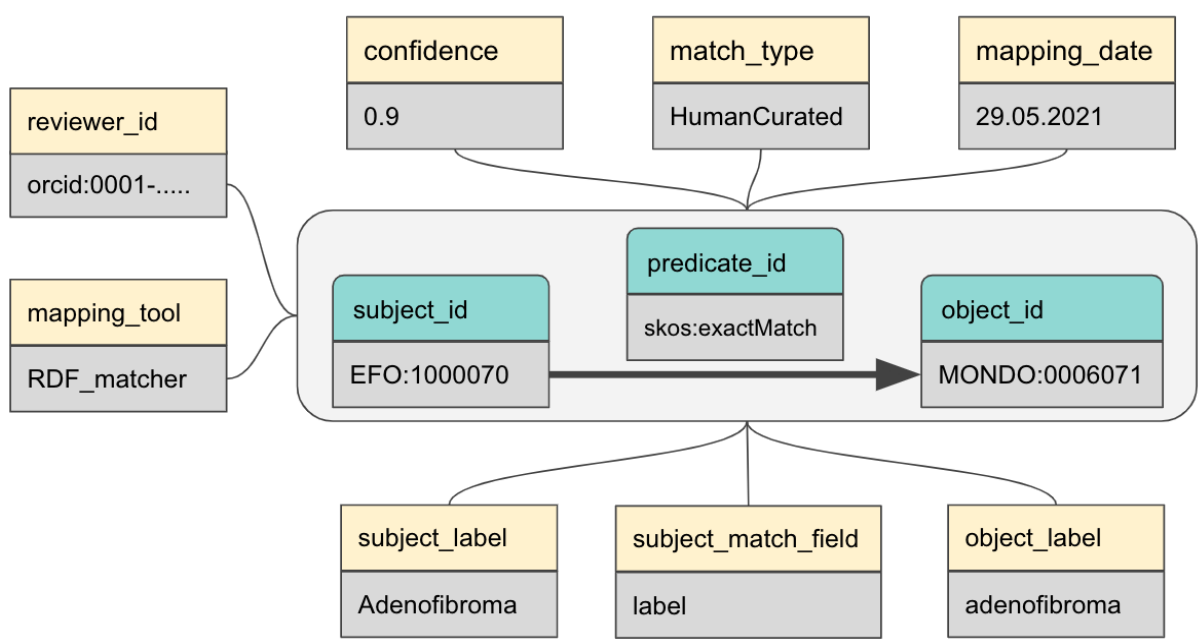

Figure 2: Example of basic SSSOM mapping model with some illustrative mapping metadata elements.

Predicates

SSSOM allows any vocabulary to be used to describe the relationship (predicate) between subject and object, but we recommend that the *predicate_id* is drawn from either SKOS or OWL vocabularies, in particular one of the predicates listed in Table 2.

| Predicate | Description |
|---|---|
| owl:sameAs | The subject and the object are instances (OWL individuals), and the two instances are the same. |
| owl:equivalentClass | The subject and the object are classes (OWL class), and the two classes are the same. |
| owl:equivalentProperty | The subject and the object are properties (OWL object, data, annotation properties), and the two properties are the same. |
| rdfs:subClassOf | The subject and the object are classes (OWL class), and the subject is a subclass of the object. |
| rdfs:subPropertyOf | The subject and the object are properties (OWL object, data, annotation properties), and the subject is a subproperty of the object. |
| skos:relatedMatch | The subject and the object are associated in some unspecified way. |
| skos:closeMatch | The subject and the object are sufficiently similar that they can be used interchangeably in some information retrieval applications. |
| skos:exactMatch | The subject and the object can, with a high degree of confidence, be used interchangeably across a wide range of information retrieval applications. |
| skos:narrowMatch | The object of the triple is a narrower concept than the subject of the triple. |
| skos:broadMatch | The object of the triple is a broader concept than the subject of the triple. |

Table 2: Recommended values of predicate_id capturing a broad range of use cases, drawn from SKOS vocabularies and from OWL.

"match_type" is a term from a controlled vocabulary that describes the method by which the match was established that led to the mapping. There are currently five types of matches in SSSOM:

1. Lexical: The match was determined through a lexical analysis of some kind
2. Logical: The match was determined by an automated reasoner (9)
3. HumanCurated: The match was determined by a human expert
4. SemanticSimilarity: The match was determined by a semantic similarity algorithm such as Resnik or Jaccard (10).
5. Complex: The match was determined by a variety of strategies, usually as part of an automated matching tool.

Each of these match types can be refined through a combination of other metadata elements. For example, a lexical match should be further qualified using the subject and object "match_field". The match field can be set to the CURIE for the property that was used to perform the match. This can be drawn from any vocabulary, but we recommend the use of properties sanctioned as part of the OBO Metadata Ontology (OMO)(11), which includes properties commonly used in OBO Foundry ontologies, including:

- rdfs:label i.e. the primary label for the matched entity
- oboInOwl:hasDbXref, for when the 'xrefs' for the entity matched
- oboInOwl:hasExactSynonym, when the match is to an exact synonym of the entity

Lastly, if the match occurred after applying preprocessing, this can be captured by the "preprocessing" metadata field. Semantic similarity matches can be further refined by providing a "semantic_similarity_score" and "semantic_similarity_measure". All automated matches, in particular complex matches, should make reference to a "mapping_tool" and its "mapping_tool_version".

Provenance

Most SSSOM metadata elements pertain to provenance. We will describe some of the most important ones here, and refer the interested reader to the full list in the online documentation (12). Mappings are maintained and established by authors ("author_id"), owned and published (i.e. brought into their SSSOM mapping form) by creators ("creator_id"), and reviewed by one or more reviewers ("reviewer_id"). For maximum transparency we recommend the use of ORCID CURIEs (13), ROR IDs (14) for organizations and Wikidata IDs (15). For example, a domain expert *(orcid:0000-0002-7356-1779)* determines that *UBERON:0002101* (limb), is an exact match to the term *FMA:24875* ("Free limb"). Domain expert *orcid:0000-0002-7356-1779* is a consultant for the European Bioinformatics Institute (EMBL-EBI, *ror:02catss52*), which produces the SSSOM mapping set that the above mapping is captured in and publishes it. Curator *orcid:0000-0002-7073-9172* reviews the mapping and confirms it.

The subject and object of a mapping each come from a source, such as an ontology or a database ("subject_source", "object_source"). For example, the term *UBERON:0002101* comes from a source "Uberon". When the mapping is created, it is usually based on a specific version of the source (e.g. "subject_source_version") which we recommend encoding with a version string such as "2020-01-01" or "2.1.0". This is important especially for making incompleteness transparent - potentially missing mappings can now be attributed to an outdated mapping set. The mapping set itself is similarly attributed an ID ("mapping_set_id") and version ("mapping_set_version").

Finally, the "mapping_date" is the date on which the mapping was established by the mapping author, and the "publication_date" is the date on which the SSSOM mapping file was published by its creator. The "why" and "what" of provenance are implicit in the model. For the "why", we expect that the intention is to map two entities in an unconditional fashion, i.e. that we model the case where the mapping is always true; see "Limitations" section. Any contextual parameters that need to be considered when interpreting the mapping should be explicit in the mapping predicate.

LinkML specification

The SSSOM schema is managed as a LinkML (16) model. LinkML, the Linked Data Modeling Language, allows schemas describing the structure of the data to be authored in YAML format. LinkML gives us a range of advantages for managing our schema:

1. We can automatically convert it into common schema representations such as JSON Schema, ShEx, SHACL, or OWL (these are all available from the GitHub repository)
2. We can use LinkML utility classes to automatically convert instance data into common representations such as JSON or RDF.
3. We can use LinkML meta models to automatically generate Python dataclasses and implement data validators etc, and we use these in our own Python toolkits (see below).

4. The SSSOM schema in YAML is easier for domain experts to read and maintain, compared to other complex schema representation languages such as OWL or JSON Schema.

## SSSOM TSV format: a simple tabular format for dissemination of mappings

A simple, table-based serialization of mapping sets was one of the core requirements for creating SSSOM. Tables are, in our experience, by far the most widely used data source in data science pipelines, and still the preferred medium for curating data. Many related approaches in the Semantic Web community such as those discussed as part of the "Semantic Web Challenge on Tabular Data to Knowledge Graph Matching" (SemTab, (17)) reflect the importance of tables as a mechanism for curating data. Therefore, SSSOM TSV format should be considered the native SSSOM data format, with other formats like RDF/XML or JSON-LD functioning as export formats. The reason a "native" format is important is that we do not only want to offer a better model for capturing metadata, but also to promote better practices for mapping as a process. One of these practices is that we want to produce mappings that are consumable and interpretable by very general toolchains, such as the ones used in data science (Pandas(18), etc). While we do provide a Python toolkit for the more advanced operations involving SSSOM mapping sets, it was a key design consideration that SSSOM mapping files should be readable by general toolkits, without the need of any special tooling.

Figure 3. An example SSSOM TSV table, with a table header (lines that start with #, shown in purple) that contains the mapping set metadata, followed by the mappings. This example is from the sssom.tsv file for ECTO, the environmental exposure ontology (19).

A SSSOM TSV table comprises two main parts: the actual table which contains the mapping and its metadata, and a table header which contains the mapping set metadata. Figure 3 shows an example of part of a simple SSSOM TSV file. The header part of the table is commented YAML (indicated by the leading # symbol). For practical purposes, we support both this "embedded" mode, where the YAML header is provided together with the mapping table, and an "external" mode, where the SSSOM YAML header is supplied as a separate file. Due to the risks involved in managing two files (losing provenance during sharing, etc.), we promote the use of the embedded mode, but the SSSOM Python toolkit can convert from external to embedded mode to ensure compatibility.

## Sustainability: Collaborative workflows and governance

The SSSOM standard is maintained as an open source project on GitHub in the mapping-commons organization. During our inaugural workshop in September 2021 (20), we established our basic governance rules (https://github.com/mapping-commons/sssom/issues/82). We make heavy use of GitHub collaborative workflows including issue templates, pull requests and reviews, GitHub actions for Continuous Integration and, perhaps most importantly perhaps, a public issue tracker to respond to and manage our interactions with the wider mapping community.

*Changing the schema.* The SSSOM schema is managed entirely as a LinkML model (16), with the source yaml file managed in GitHub. To change the schema, we perform the following actions. For every change (usually adding/changing metadata elements), we require the creation of a GitHub issue detailing the nature of the change. This ensures that the community has time to respond to the intended change even before it is performed. If the community reaches agreement on the nature of the change, an edit to the source schema is created and a GitHub pull request is opened. The pull request stays open for review. If the schema change is not backwards compatible (i.e. current SSSOM mappings are affected), the change needs to be approved by members of the core team.

## The SSSOM software ecosystem

There are several useful tools for working with SSSOM. *sssom-py* is a Python library and a command-line toolkit that was designed to work with SSSOM (21). The library covers functionality such as importing files from different formats (OBO Graphs JSON (22), RDF Alignment API (23)) and exporting them as SSSOM tables; converting SSSOM tables to RDF, OWL or JSON-LD; merging and querying SSSOM tables and validating them. For an overview of the full functionality of sssom-py, refer to the documentation (24).

*rdf-matcher* is a matcher for RDF vocabularies or OWL ontologies that exports mapping sets as SSSOM tables, including mapping rules (25). For example, rdf-matcher exports metadata such as mapping tool, confidence, match fields and match string. It can document simple mapping rules such as matches on label and synonym fields.

In our vision for the publication of terminological mappings, related mapping sets are collected and even maintained as part of a *mapping commons*. A mapping commons is a public registry that enables users to find mappings for a clearly defined use case such as "cross-species phenotype mappings" or "disease mappings". An example of a mapping commons that focuses on mappings related to mice and humans can be found on GitHub (26). The creation of mapping commons is in the very early stages, but the hope is that users can simply report wrong mappings much the same way as they can document issues on other semantic artefacts such as ontologies or terminologies.

## Why we need better metadata for terminological mappings: Use cases

Here we describe four use cases that motivated the development of SSSOM.

## Use Case 1: Harmonizing disease mappings: Mondo Disease Ontology

The Mondo disease ontology (27) seeks to harmonize a variety of disease ontologies and terminologies in a consistent logical framework. Mondo not only provides semantically precise mappings to external sources; it also ensures that these mappings are reconciled, i.e. no single external term will ever map to more than one term in Mondo. This enables users to map their disease data to Mondo from a variety of sources such as Online Mendelian Inheritance in Man (OMIM, (28)), Disease Ontology (DO, (29)), Orphanet (30), National Cancer Institute Taxonomy (NCIT, (31)) and the International Classification of Diseases (ICD, (32)), and analyze the data in a coherent logical framework.

Maintaining a harmonized set of mappings is a complex task. To make the integration of more terminological sources and the ongoing maintenance of mappings scalable, Mondo uses an automated Bayesian approach for ontology merging (k-BOOM, (33)) which takes as an input the two ontologies to be aligned and a set of mappings with probabilities. These mapping sets can be generated by any matching tool, as long as there is some kind of confidence/probability score and a precise mapping predicate (e.g., *skos:exactMatch*, *skos:narrowMatch*, etc.). The current implementation of the k-BOOM algorithm in the Boomer tool (34) reads SSSOM files as mapping candidates and then discovers the most likely "correct" mappings. Tools like Boomer rely on mappings with transparent imprecision and accuracy to work effectively.

In addition to maintaining a set of harmonized mappings, Mondo also has to distribute them. Before SSSOM, mappings were primarily distributed as owl:equivalentClass axioms and skos:exactMatch (or even *oboInOwl:hasDbXref*) annotations, which made them hard to use for any but ardent users of semantic web technologies. Mondo now exports SSSOM tables as part of their release pipeline. Because these tables include explicit provenance information, they allow downstream users to use the mappings effectively.

## Use Case 2: Browsing and cross-walking mappings: OxO

The European Bioinformatics Institute (EBI) developed the Ontology Xref (Cross-reference) Service (OxO, (35)) to enable users to find suitable mappings for their ontology terms and provide APIs to access them (36). OxO integrates cross-references from OBO ontologies, and mappings from UMLS and other sources. Users make heavy use of OxO's ability to "walk" mappings. "Walking" (also known as cross-walking or hopping) is the ability to link terms together based on intermediate mappings. For example, a user might look for suitable mappings for *FMA:24875* ("Free limb"), e.g. https://www.ebi.ac.uk/spot/oxo/terms/FMA:24875. Within mapping distance 1 (1 hop) we only find a single suitable match at the time of this writing (October 2021), *UBERON:0002101* ("limb"). If we increase the search radius to mapping distance 2, we find 7 additional mappings which look fine, like *MA:0000007* (limb) or *NCIT:C12429* (Limb). However, we also see the first issues emerge: *EFO:0000876* (obsolete vertebrate limb) and *UMLS:C0015385* (Extremities) are also among the search results. Terms that are marked as obsolete should not be used in mappings, and the term "extremities" usually refers to appendages such as hands or feet rather than the whole limb. Indeed, on closer inspection, we find that "Extremities" is mapped to *UBERON:0000026* (appendage) in OxO. A blind application of walks cannot work if we do not know that the mapping from "limb" to "extremities" is related rather than exact - rather than being simply "cross-references" without precise semantics, we need our mappings to be transparent about imprecision. OxO was designed as a tool to query and walk "cross-references" which do not have "precision" by design - they often correspond to exact matches, but they can correspond to broad, narrow, close or related matches, without explicitly specifying that as part of the metadata. This captures the original use case of OxO: finding closely related terms across terminologies and ontologies. With the advent of SSSOM, OxO seeks to enable further use cases, like cross-walks with precise mappings, by capturing additional metadata. A first draft of this extension to the current OxO data model and a prototype user interface is planned for May 2022.

Users want to be able to view only the trustworthy mappings, and what we deem "trustworthy" is very much dependent on our personal experience and preference. While it is already possible to restrict search in OxO to particular sources, OxO imports all cross-references found in these sources, disregarding any additional metadata. For example, unlike Mondo (described in Use Case 1), OxO currently does not distinguish between *skos:exactMatch* and *skos:relatedMatch*. To convince users that a particular mapping is good enough for their particular use case, we may need to present the mapping rules that were applied to determine the mappings. Such metadata does not currently exist at all in most mapping sets, but in order to curate and then leverage it, we must first provide standards like SSSOM to represent common mapping rules, which can then be implemented by tools like OxO.

## Use Case 3: National Microbiome Data Collaborative

The National Microbiome Data Collaborative (NMDC, (37)) integrates environmental omics-related data and metadata from multiple sources. This involves aligning metadata schemas from multiple different sources including the Genomes OnLine Database (GOLD), NCBI, the NMDC schema, and the Genomics Standards Consortium Minimum Information about any (x) Sequence (MIxS (38)) standard. It also involves aligning underlying vocabularies used to describe categorical aspects of samples, including the GOLD environmental path vocabulary and the Environment Ontology (ENVO) (39).

NMDC has created SSSOM files for these mappings, making use of multiple aspects of the SSSOM standard, including the mapping predicate (most mappings are exact, but a small handful are related matches), and whether the mapping has been curated by an expert or was obtained from a specific source. Using SSSOM allows the NMDC to use standard tools for summarizing and validating these mappings.

## Use Case 4: Finding and using mappings in EOSC-Life

The EOSC-Life project (https://www.eosc-life.eu/) brings together the 13 Biological and Medical ESFRI research infrastructures to create an open collaborative space for digital biology in Europe. EOSC-Life has designed a use case with Alice as a data steward who needs to register patient information in the European registry for Osteogenesis imperfecta, which uses Orphacodes (from Orphanet) for diseases, whereas other partners use SNOMED CT. To demonstrate the automatic conversion from SNOMED CT to Orphacodes, they set up a FAIR Data Point (FDP) with the metadata description of the mappings and used SSSOM to describe mappings. FDP is a realization of FAIR data principles that stores database metadata and publishes it on the web. Compared with the simple equivalence between the objects of the same subject provided by other mapping systems, mappings described by SSSOM have richer metadata, e.g. specifying match precision. Combining FDP and SSSOM, it is possible for Alice to access mappings via a FDP according to their semantics, and to automatically use the mappings by converting specified subjects to the mapped objects accurately via SSSOM metadata.

# Related work

In this section, we discuss alternative formats for capturing terminological mappings, as well as some less formal efforts concerned with mapping metadata. We also describe some of the impactful mapping-related tools and discuss how they could benefit from implementing a standard mapping metadata model such as SSSOM.

## Standard formats for capturing mappings

The RDF Alignment Format (40) is currently the main format used to exchange mappings within the ontology matching community and the Ontology Alignment Evaluation Initiative (OAEI, (41), (42), (43)), a coordinated international initiative to forge consensus for evaluation of ontology matching methods. The main advantage of the RDF Alignment format is its simplicity. EDOAL (44) is a more expressive format that aims at representing complex mapping, e.g. linking two or more entities beyond atomic subsumption and equivalence. Both the RDF Alignment and EDOAL formats are supported by the Alignment API (45). SSSOM, like the RDF Alignment Format, brings a simple format to exchange mappings, which improves on the metadata and provenance information associated with the mappings, enhancing their understanding and potential future reuse. The Matching Evaluation Toolkit (MELT, (46)) is a framework for developing, tuning, evaluating, and packaging ontology matching systems that has been adopted by OAEI. Currently MELT supports the RDF Alignment and EDOAL formats, but it is a modular framework that can easily support additional mapping exchange formats like SSSOM. The Ontology Matching community is considering adopting SSSOM as an additional mapping exchange format.

The Vocabulary of Interlinked Datasets (VoID, (47)) is a W3C Interest Group RDF Schema vocabulary for expressing metadata about RDF datasets (48). Beyond describing RDF datasets in general, VoID allows the specification of Linksets, i.e. collections of links where the subject is in a different dataset than the object. VoID metadata elements are fairly high level and need to be extended to capture fine grained provenance and mapping rules. The Open PHACTS project extended VoID, in particular, to support mapping justifications using the BridgeDb Mapping Vocabulary (49). BridgeDb (50)) is an open source data identifier mapping service that is typically used for mappings between genes and gene products, metabolites and reactions. In principle BridgeDb can also provide ontology mappings; for instance, it has already been used for gene-disease relationships. BridgeDb can stack mappings; the most common use case for that is when people use their own ontology or identifier class and want to map these first to an external identifier and then to relate them using standard mappings. A semantic web version of BridgeDb was developed as the OpenPHACTS Identifier Mapping Service (51). We are discussing integrating VoID linksets and the BridgeDb mapping vocabulary with SSSOM mappings mapping sets, which provide richer metadata. However, some obstacles exist. For example, the VoID specification does not permit multiple subjects or objects in a single linkset file, which is a critical requirement of SSSOM.

Some mapping metadata can be captured using simple established vocabularies such as Dublin Core (52) and OWL (3). There are also a variety of approaches to capturing more detailed provenance, such as PROV-O (https://www.w3.org/TR/prov-o/), PAV (https://pav-ontology.github.io/pav/), and the Mapping Quality Vocabulary (MQV) (https://alex-randles.github.io/MQV/). The SSSOM data model allows some basic provenance information to be captured using properties such as *creator_id* and *mapping_provider.* Currently, two SSSOM properties are mapped directly to PAV and one to PROV-O but work is underway to provide a complete mapping to these established standard vocabularies, including a more comprehensive alignment with the PROV-O activity model.

The Distributed Ontology, Modeling, and Specification Language (DOL, (53)) is an Object Management Group (OMG, (54)) standard for the representation of distributed knowledge, system specification and model-driven development across multiple ontologies, specifications and models (OMS). DOL enables the representation of alignments across OMS that have been formalized in different formal (logical) languages on a sound and formal semantic basis. In contrast to SSSOM, DOL deals primarily with distributed semantics and does not define a vocabulary for mapping metadata and mapping rules. Another OMG standard that includes a component for terminological mappings is the Common Terminology Services 2 System (CTS2, (55)). CTS2 supports the management, maintenance, and interaction with ontologies and medical vocabulary systems, providing a standard service information and computational model. The CTS2 Map Services specify how entity references from one code system or value set are mapped to another. The CTS2 map entry information model reflects many of the metadata elements also defined by SSSOM, such as subject and object source references, version information, and mapping set names. In contrast to SSSOM, CTS2 allows mapping one entity to multiple targets in complex mapping expressions and specifies the expected behavior of mapping services. Overall, it is considerably more complex than SSSOM and geared towards interoperability between software systems in a wider clinical context rather than FAIR exchange of terminological mapping.

The Systematized Nomenclature of Medicine -- Clinical Terms (SNOMED CT) is a clinical terminology designed to represent content in electronic health records (56). The SNOMED CT logical model, unlike SSSOM, does not require extensive provenance on how a mapping was created. Another difference between these resources is transparency and accessibility. Currently, the mappings provided by SNOMED CT must be built locally using their OTF-Mapping-Service (https://github.com/IHTSDO/OTF-Mapping-Service). SSSOM is a fundamental component underlying all of the Mapping Commons (https://github.com/mapping-commons), which means all projects within the Commons are interoperable and publicly available. Perhaps the most important distinction is that unlike the SNOMED CT Reference Sets, which are explicitly designed for use with SNOMED-specific resources, SSSOM is not designed for use with a single system, infrastructure or standard.

The Unified Medical Language System (UMLS) (57) is a repository of biomedical vocabularies developed by the US National Library of Medicine (NLM). The NLM coordinates a number of mapping efforts, such as SNOMED to ICD 10 (58). UMLS maps 214 vocabularies based on automated approaches that exploit lexical and semantic processing and manual curation (59). The UMLS API can exploit mappings to enable cross-walks, much the same way as OxO does (see Section on Use Cases). While the UMLS mapping model is probably the closest to a standard tabular representation for mapping metadata, it lacks many of the metadata elements defined by SSSOM, and, more importantly, does not define a public, collaborative workflow for defining new metadata elements or formally defining mappings into other formats such as RDF or JSON.

## Informal approaches for capturing mappings

In addition to the mapping standards described above, there have been various less formal attempts to capture mappings. Some of these were launched to address a specific need but fail to address some of the requirements that SSSOM satisfies. Many ontologies in OBO make use of the *oboInOwl:hasDbXref* property (60), also known as the "database cross reference", as historically most mappings have been created as simple cross-references in this fashion. A drawback of these "xref" mappings is that they are semantically ambiguous and are used in vastly different ways in different ontologies (61). After analyzing about a million such database cross references across OBO ontologies, Laadhar et al. concluded that their unclear semantics makes them "impractical or even impossible to reuse" - a viewpoint that the authors of this paper share.

For the BioHackathon 2015, members of the DisGeNET team (62) and their collaborators surveyed a number of sources to define a minimal set of attributes and standards for ontology mapping metadata (63). The BioHackathon 2015 never resulted in a formal specification for mapping metadata, but we are now working with the DisGeNET team to incorporate the metadata elements of their survey directly into SSSOM. Most of their proposed metadata elements have already been mapped to SSSOM; others are being currently revised.

The Semantic Mapping Framework (SEMAF) is a European Commission funded project designed to create, document, and publish FAIR mappings between artefacts, e.g. vocabularies, ontologies, and lexicons, used in multiple scientific domains (64). To understand the limitations of existing mapping approaches and develop reasonable solutions, the SEMAF Task Force conducted interviews with 25 experts from a wide range of bioscientific domains and reviewed 75 reports on existing research infrastructure (65). From this work, the SEMAF Task Force identified an extensive set of requirements that span infrastructure, architecture, data models, user interfaces, machine access, optional and content management, and implementation. Their framework, which consists of a Federative Registry and a Mapping Model, was designed to support these requirements. There are many aspects of the SEMAF Mapping Model that align with the Mapping Commons principles and with SSSOM. Both SSSOM and SEMAF are built on Semantic Web principles and use assertions to provide additional metadata about a mapping, e.g. mapping provider, creation date. As the SEMAF model is still emerging, it is not yet clear how exactly it will be implemented (schemas, toolkits) and which metadata elements will be included. It is, however, our understanding from the current documentation that SSSOM would be a suitable implementation for the abstract SEMAF mapping model.

OMOP2OBO (https://github.com/callahantiff/OMOP2OBO) is the first health system-wide semantic integration and alignment between the Observational Health Data Sciences and Informatics' Observational Medical Outcomes Partnership (OMOP) standardized clinical terminologies and OBO biomedical ontologies (66). The OMOP2OBO framework provides both a mapping algorithm and an open source repository of mappings. OMOP2OBO uses a sophisticated mechanism for converting flat-file mappings into RDF and OWL (67), which is currently being aligned with SSSOM.

## Mapping services and tools

The EMBL-EBI Ontology Xref Service (OxO) (https://www.ebi.ac.uk/spot/oxo/) provides a Web-based user interface and REST API to allow retrieval of mappings between terms. We are working with the OxO team to extend their mapping model to support SSSOM natively, (see section on use cases). This will make the output of OxO more useful by including further information about mappings, and allow OxO to ingest mappings from any SSSOM datasource.

The BioPortal software (along with its deployed versions based on the OntoPortal distribution) manages mappings of multiple types from a variety of sources ((68), (69)), and presents them in two contexts (ontology to ontology, and term-to-term) and via two access methods (UI and API). Types of mappings include URI (same IRI in both places), CUI (matching CUI values in UMLS terms), LOOM (a syntactical match using the LOOM algorithm, (70)), and REST (mappings provided by BioPortal users). BioPortal automatically creates the URI, CUI, and LOOM mapping information each time an ontology is updated. Several metadata attributes are stored with each mapping, including a timestamp, the process that created the mapping (including the user who provided each REST mapping), the mapping relationship and mapping type. As with OxO, a user submitting REST mappings to BioPortal must convert their data to the defined BioPortal mapping submission format. The BioPortal team intends to provide SSSOM support and a SPARQL endpoint as part of a new NIH grant, and to provide this code as part of its shared OntoPortal Appliance distribution, used by repositories such as AgroPortal and EcoPortal.

Biomappings (71) is a repository for both curated and predicted mappings along with their associated metadata. It is intended to fill in the gaps in the availability of mappings between widely used resources. Biomappings is built on public tools such as git and GitHub, uses automated testing and continuous integration to check data integrity, provides a web-based curation interface for triaging predicted mappings and adding novel ones, and offers several workflow examples for generating new predictions using Gilda (72) or custom scripts. Its data are available under the CC0 1.0 license and distributed in the SSSOM format to promote contributions, reuse, and enable incorporation into primary resources.

The Pistoia Alliance best practice guidelines were designed to check how suitable source ontologies are for mapping (73). They emphasize the application of ontologies in the life sciences to encourage best practices and to aid mapping of ontologies in a particular domain. This public resource was developed as part of the Pistoia Alliance Ontologies Mapping project, which also defined the requirements for an Ontologies Mapping tool and service (74), (75). This led to the development of Paxo (76), a lightweight Ontology Mapping tool designed to align ontologies hosted by the OLS (77) and to integrate them with OxO (36). The alignments generated from Paxo and OxO are available, but they are currently limited to CSV format. These mapping alignments would benefit greatly from transformation into the much more expressive SSSOM format to capture the relevant metadata, and some of them have already been converted (https://github.com/mapping-commons/mh_mapping_initiative).

## Discussion and limitations of the approach

In this section, we discuss shortcomings of the current SSSOM approach, and potential ways to address them:

1. Mappings themselves have no context (i.e. are always true)
2. Complex mapping rules are hard to capture due to the simple, flat data model
3. Mappings are not idempotent, i.e. there are metadata elements that modify one another
4. Lack of support for complex mappings

***Mappings have no (global) context.*** There are many mapping scenarios, especially in the clinical domain, where mappings only hold under a range of applicability criteria. For example, we could say that "*UBERON:0002101* (metazoan limb) is equivalent to *XAO:0003027* (xenopus limb) under the assumption that taxon constraints are ignored. Or we might want to express that you can swap one term from a clinical terminology for another, but only if we can assume the patient is an adult female. It was an important design decision for SSSOM to decide that mappings should be universally applicable and not dependent on some global context, which would make merging and reconciling them much more complex (requiring specialized tooling). While this can be a significant problem for some use cases, there are two potential workarounds (one that is currently supported, and one that is currently under discussion): (1) the contextual parameters of the mapping can be captured as part of the mapping relation. For example, one could define a new relation "example:hasExactCrossSpeciesMatch" as a sub-relation of skos:closeMatch that links *UBERON:0002101* and *XAO:0003027*. The problem with this approach is, while currently supported, it would push the contextual parameters far away into an ontology of relations, which mapping applications, for example, would have to import and interpret. (2) The contextual parameters could be captured as complex expressions. For example, you could define *UBERON:0002101* (limb) + *NCBITaxon:8353* (xenopus) → *XAO:0003027* (xenopus limb), and capture "*UBERON:0002101* (limb) + *NCBITaxon:8353* (xenopus)" as an ontological class expression such as "*UBERON:0002101* and 'in-taxon' some *NCBITaxon:8353*". This is currently not supported, but is being discussed. Ultimately the balance here is between capturing all mapping scenarios and keeping the metadata and format as simple as possible. It seems to be the case that a large percentage of use cases can be captured without introducing complex expressions.

***Complex mapping rules are hard to capture in a simple, flat model.*** Many matching decisions, in particular those done by automated tools, are complex: they involve a variety of mapping rules. For example, an automated matching tool may determine that based on a specific threshold of semantic similarity, e.g. > 0.9, and a matching label, we decide that subject-predicate-object triple constitutes a match. A flat data model like SSSOM cannot easily capture the case where a match is associated with multiple match types. Again, this modeling decision comes down to the simplicity vs. expressivity tradeoff described above. While it would be easy to build a data model that supports multiple complex match types, it violates one of our central requirements: being able to express the data as a simple table. We therefore decided to accept this shortcoming. For our use cases that require complex match types, we therefore provide "one row per mapping rule". Our reference implementation, rdf-matcher, for example, would produce two rows for the match between *UBERON:0002101* and *FMA:24875* if there was a lexical match on an exact synonym of the terms and also a lexical match on the primary labels of the terms.

***Mappings are not idempotent: adding a column to a mapping table could change its semantics.*** The hardest design decision to make was regarding the modifiers on "predicate_id". There are many use cases for modifiers, such as negation: you want to be able to say that *UBERON:0002101* is NOT a skos:exactMatch to *FMA:54448*. After debates during our 1st Workshop on SSSOM (20), we decided to add a "predicate_modifier" element to SSSOM which allows such encodings. The alternative would have been to introduce additional syntax (e.g. !*skos:exactMatch*) or additional predicates like "example:notExactMatch". The former solution (!*skos:exactMatch*) is a violation of the "simplicity" requirement because it introduces the need to handle special syntax on the user side. The latter solution would have led to a potential doubling of all predicates - which could have led to a combinatorial explosion if it had to capture additional modifiers, such as "direct" or "inverse". The main limitation, and risk, of our chosen approach is that users that consume SSSOM may simply believe that the mappings they consume do not have a predicate modifier (because they never had in the past), and therefore not notice that they suddenly consume "negative" or otherwise modified mappings. We decided that this risk was worth it to keep the model simple and easy to use. A second example where we violate idempotency, i.e. where the addition of additional metadata could change the semantics of pre-existing metadata, is with our "preprocessing" fields - ignoring the "preprocessing" field when interpreting the "match_field" columns could lead to confusing results. For example, "Alzheimer 2" and "Alzheimer 3" are different concepts, but if the non-alphabetical characters were stripped during pre-processing, they would be (misleadingly) matched.

***Lack of support for complex mappings.*** Complex mappings are currently not supported by SSSOM. A complex mapping is a mapping where at least one of the subject or objects of a mapping does not correspond directly to a term, but rather to an expression involving more than one term. Complex mappings are hard to evaluate, but they are receiving interest in the literature (78). During our first workshop, we discussed how complex mappings can be represented, but the majority of the participants were in favor of postponing the introduction of complex mapping to a later stage to protect the simplicity of the current metadata model. We are considering implementing an extension for SSSOM that can capture complex mappings.

# Conclusions and Future Work

Despite the importance of mappings for data integration, standardizing the representation of mappings and mapping rules has not received the same level of care as other "semantic artefacts" such as controlled vocabularies and ontologies. For many use cases, merely providing the subject, object and even predicate of a mapping is not enough, and many mapping sets suffer from non-transparent imprecision, non-transparent inaccuracy, non-transparent incompleteness and unFAIRness, in particular in terms of re-usability. Attempts to standardize the representation of mappings are scarce, and generally fall short in three important areas:

1. Insufficient vocabulary for describing metadata in a way that make imprecision, inaccuracy and incompleteness explicit.
2. Lack of free, open and community-driven collaborative workflows that are designed to evolve the standard continuously in the face of changing requirements and mapping practices.
3. Lack of a standardized tabular representation of a mapping set, which is imperative for facilitating both human curation and use in data science pipelines, and integrates seamlessly with the Linked Data stack.

SSSOM addresses these shortcomings by providing a rich vocabulary for describing mapping metadata, being entirely community-driven with sustainable governance processes in place, and promoting a very simple tabular format for the dissemination of mappings that can be easily integrated in typical data science workflows.

Having a simple standardized format is the main prerequisite for generating high quality mappings and facilitating their sharing and reuse. The next step, however, is probably even more important, and more difficult: establishing shared best practices for building better mappings. The authors have been working with various groups on improving their manual and automated mapping practices. For example, we worked with OpenTargets (79) to disseminate Mondo-Meddra mappings in SSSOM format. A simple standard for mapping metadata and a simple table format have been instrumental for collaborating with groups such as OpenTargets, IMPC (80), MGI (81) and the Center for Cancer Data Harmonization (CCDH, (82)) to build better mapping sets, for example by sharing and editing mapping sets directly through Google Sheets. Developing a metadata standard is usually not enough to improve the quality of data (in our case, mappings) and needs to be accompanied by a set of shared best practices. Analogously to the 5-Star deployment scheme for Linked Data developed by Sir Tim Berners-Lee ([https://www.w3.org/2011/gld/wiki/5_Star_Linked_Data](https://www.w3.org/2011/gld/wiki/5_Star_Linked_Data)), we are developing a 5-Star scheme for rating mappings (83). This 5-star scheme directly evolved out of our experiences working with our collaborators to understand how to best document mappings. It includes considerations such as where and how mapping sets should be published, how they should be licensed and which metadata should be provided.

Our focus in the near future is on developing training materials to help groups to build better mappings, while continuing to evolve the SSSOM standard and the associated Python toolkit and extending the OxO mapping repository to support SSSOM. We appreciate that not all mapping use cases can be captured by a representation that is deliberately simple, but hope that the medical terminology, database and ontology mapping communities will embrace this more principled approach to disseminating FAIRer and more reusable mapping sets.

## Acknowledgements

Some of the authors, including NM and CJM, were partially supported by R24-OD011883 (Office of the Director, National Institutes of Health). QL has received funding from the European Union's Horizon 2020 research and innovation programme under grant agreement No 824087 (EOSC-Life). This work was supported in part by the Director, Office of Science, Office of Basic Energy Sciences, of the US Department of Energy (Contract No. DE-AC0205CH11231). CTH and BMG were supported by the DARPA Young Faculty Award W911NF2010255. H. Harmse and SJ have received funding from EOSC-Life from the European Union's Horizon 2020 programme under grant agreement number 824087 and EJP-RD Initiative funding from the European Union's Horizon 2020 research and innovation programme under grant agreement N°825575 and EMBL-EBI Core Funds
JAM and SJ have received funding from the Chan-Zuckerberg Initiative award for the Human Cell Atlas Data Coordination Platform to EMBL-EBI and EMBl-EBI Core Funds. We thank the Open PHACTS partners and funders (IMI-JU grant no. 115191) and the BioHackathon 2015 organizers for their contributions. We acknowledge the support of a gift from Bosch corporation to support this work.